\documentclass{iaus}
\usepackage{graphicx}

\title[Solar abundances and 3D model atmospheres]{Solar abundances and 3D model atmospheres}

\author[Ludwig, Caffau, Steffen, Bonifacio, Freytag, \&\ Cayrel]   
{Hans-G{\"u}nter Ludwig$^{1,2}$, 
Elisabetta~Caffau$^2$,
Matthias~Steffen$^3$,
Piercarlo~Bonifacio$^{1,2,4}$,
Bernd~Freytag$^{1,2,5}$,
\and 
Roger~Cayrel$^2$}

\affiliation{%
$^1$CIFIST -- Marie Curie Excellence Team\\
$^2$GEPI -- Observatoire de Paris,  CNRS, Universit{\'e} Paris Diderot, 92195 Meudon, France\\
$^3$Astrophysikalisches Institut Potsdam, An der Sternwarte 16, 14482 Potsdam, Germany\\
$^4$INAF -- Osservatorio Astronomico di Trieste, via Tiepolo 11, 34143 Trieste, Italy\\
$^5$CRAL --  UMR 5574 CNRS, Universit{\'e} de Lyon, {\'E}cole Normale Sup{\'e}rieure de Lyon,
46~all{\'e}e d'Italie, 69364~Lyon Cedex~07, France\\
}

\pubyear{2009}
\volume{265}  
\pagerange{119--126}
\jname{Chemical Abundances in the Universe: Connecting First Stars to Planets}
\editors{K. Cunha, M. Spite \& B. Barbuy, eds.}

\newcommand{\COBOLD}{{\sf CO$^5$BOLD}}

\newcommand{\beq}{\begin{equation}}
\newcommand{\eeq}{\end{equation}}

\begin{document}

\maketitle

\begin{abstract}
  We present solar photospheric abundances for 12 elements from optical and
  near-infrared spectroscopy. The abundance analysis was conducted employing
  3D hydrodynamical (\COBOLD) as well as standard 1D hydrostatic model
  atmospheres. We compare our results to others with emphasis on discrepancies
  and still lingering problems, in particular exemplified by the pivotal
  abundance of oxygen. We argue that the thermal structure of the lower solar
  photosphere is very well represented by our 3D model. We obtain an excellent
  match of the observed center-to-limb variation of the line-blanketed
  continuum intensity, also at wavelengths shortward of the Balmer jump..
\end{abstract}

\firstsection 

\section{Motivation}

In recent years several solar abundances experienced a significant downward
revision, among them major contributors to the overall solar metalicity
(\cite[Asplund et al. 2005]{AGS05}). In part, the downward revision was attributed to the
application of 3D model atmospheres. Due to the importance of the solar
composition as a fundamental ``yardstick'' in astronomy, the
CIFIST\footnote{Cosmological Impact of the FIrst STars, an EU funded Marie
  Curie Excellence project} Team and its collaborators started an independent
investigation of the solar abundances applying its self-developed analysis
tools, in particular its own 3D model atmosphere code dubbed \COBOLD\
(\cite[Freytag et al. 2002]{freytag02}, \cite[Wedemeyer et al
2004]{wedemeyer04}).  Table~\ref{t1} summarizes the result for 12 elements in
comparison to other works. Considering the latest compilation of Asplund and
collaborators one can note a convergence towards a unique abundance set.
However, there are still sizable differences present, in particular concerning
the abundant element oxygen.  Formally, the overlapping error bars could be
taken to basically signal correspondence. However, one must keep in mind that
certain systematics (observed spectra, oscillator strength, analysis
methodology) are shared among all groups, and from that perspective
differences are still on a rather high level. In this contribution we want to
comment on a few of the lingering problems when it comes to the spectroscopic
determination of solar abundances.

\begin{table}
\begin{center}
\caption{Abundances derived by the \COBOLD\ group in comparison
  to other compilations: AG89 \cite{AG89}; GS98 \cite{GS98}; AGS05
  \cite{AGS05}; AGSS09 \cite{AGSS09}. El denotes the element, N the number of
  spectral lines used in our analysis. The last two rows give the total mass
  fraction of metals, and metals relative to hydrogen. Values set
  in italics refer to meteoritic abundances.}\label{t1}
{\small
\begin{tabular}{|l|r|l|l|l|l|l|}
\hline
El & N  & \COBOLD & AG89 & GS98 & AGS05 & AGSS09 \\
\hline
Li & 1  & $1.03\pm 0.03$ & $1.16\pm 0.10$ & $1.10\pm 0.10$        & $1.05\pm 0.10$        & $1.05\pm 0.10$ \\
C  & 43 & $8.50\pm 0.06$ & $8.56\pm 0.04$ & $8.52\pm 0.06$        & $8.39\pm 0.05$        & $8.43\pm 0.05$ \\
N  & 12 & $7.86\pm 0.12$ & $8.05\pm 0.04$ & $7.92\pm 0.06$        & $7.78\pm 0.06$        & $7.83\pm 0.05$ \\
O  & 10 & $8.76\pm 0.07$ & $8.93\pm 0.035$& $8.83\pm 0.06$        & $8.66\pm 0.05$        & $8.69\pm 0.05$ \\
P  & 5  & $5.46\pm 0.04$ & $5.45\pm 0.04$ & $5.45\pm 0.04$        & $5.36\pm 0.04$        & $5.41\pm 0.03$ \\
S  & 9  & $7.16\pm 0.05$ & $7.21\pm 0.06$ & $7.33\pm 0.11$        & $7.14\pm 0.05$        & $7.12\pm 0.03$ \\
Eu & 5  & $0.52\pm 0.03$ & $0.51\pm 0.08$ & $0.51\pm 0.08$        & $0.52\pm 0.06$        & $0.52\pm 0.04$ \\
Hf & 4  & $0.87\pm 0.04$ & $0.88\pm 0.08$ & $0.88\pm 0.08$        & $0.88\pm 0.08$        & $0.85\pm 0.04$ \\
Th & 1  & $0.08\pm 0.03$ & $0.12\pm 0.06$ & ${\it 0.09\pm 0.02}$  & ${\it 0.06\pm 0.05}$  & $0.02\pm 0.10$ \\
K  & 6  & $5.11\pm 0.09$ & $5.12\pm 0.13$ & $5.12\pm 0.13$        & $5.08\pm 0.07$        & $5.03\pm 0.09$ \\
Fe & 15 & $7.52\pm 0.06$ & $7.67\pm 0.03$ & $7.50\pm 0.05$        & $7.45\pm 0.05$        & $7.50\pm 0.04$ \\
Os & 3  & $1.36\pm 0.19$ & $1.45\pm 0.10$ & $1.45\pm 0.10$        & $1.45\pm 0.10$        & $1.25\pm 0.07$ \\
   &    &                         &                &                       &                       & \\
Z  &    & 0.0153         & 0.0189         & 0.0171                & 0.0122                & 0.0134 \\
Z/X&    & 0.0209         & 0.0267         & 0.0234                & 0.0165                & 0.0183 \\
\hline                                            
\end{tabular}
} 
\end{center}
\end{table}

\section{Sources of systematic uncertainties}

While it may appear straight-forward to conduct a spectroscopic abundance
determination there are a number of sources of systematic uncertainties which
we list in the following. We comment on two selected aspects in more detail in
subsequent sections. i) Are the selected lines appropriate? The issue of
blending is an important and often difficult aspect to judge. The accuracy of
atomic data is evidently also fundamental. ii) How accurate are our
measurements of the lines' equivalent width? The ever-lasting problem of the
continuum placement constitutes a difficult to overcome limit to the
achievable precision. Line shapes fitted to observations can mitigate but not
eliminate this precision bottleneck. iii) How good are our model atmospheres,
in particular 3D models? There have been long-lasting arguments about
insufficient spatial resolution, and wavelength resolution when representing
the energy exchange between gas and radiation field. iv) How great are the
departures from local thermodynamic equilibrium? In particular, the poorly
constraint efficiency of collisions with neutral hydrogen atoms in the
calculation of the statistical equilibrium established a limit to which one
can determine abundances from some lines. Prominent examples are the infrared
triplet lines of neutral oxygen. v) Which solar spectrum is {\it the\/} solar
spectrum? There are surprising differences among high quality solar atlases
which need to be better understood -- or even better overcome by a newer
generation of atlases.

\section{3D model properties}

\begin{figure}[t]
\begin{center}
 \includegraphics[height=0.92\textwidth,angle=90]{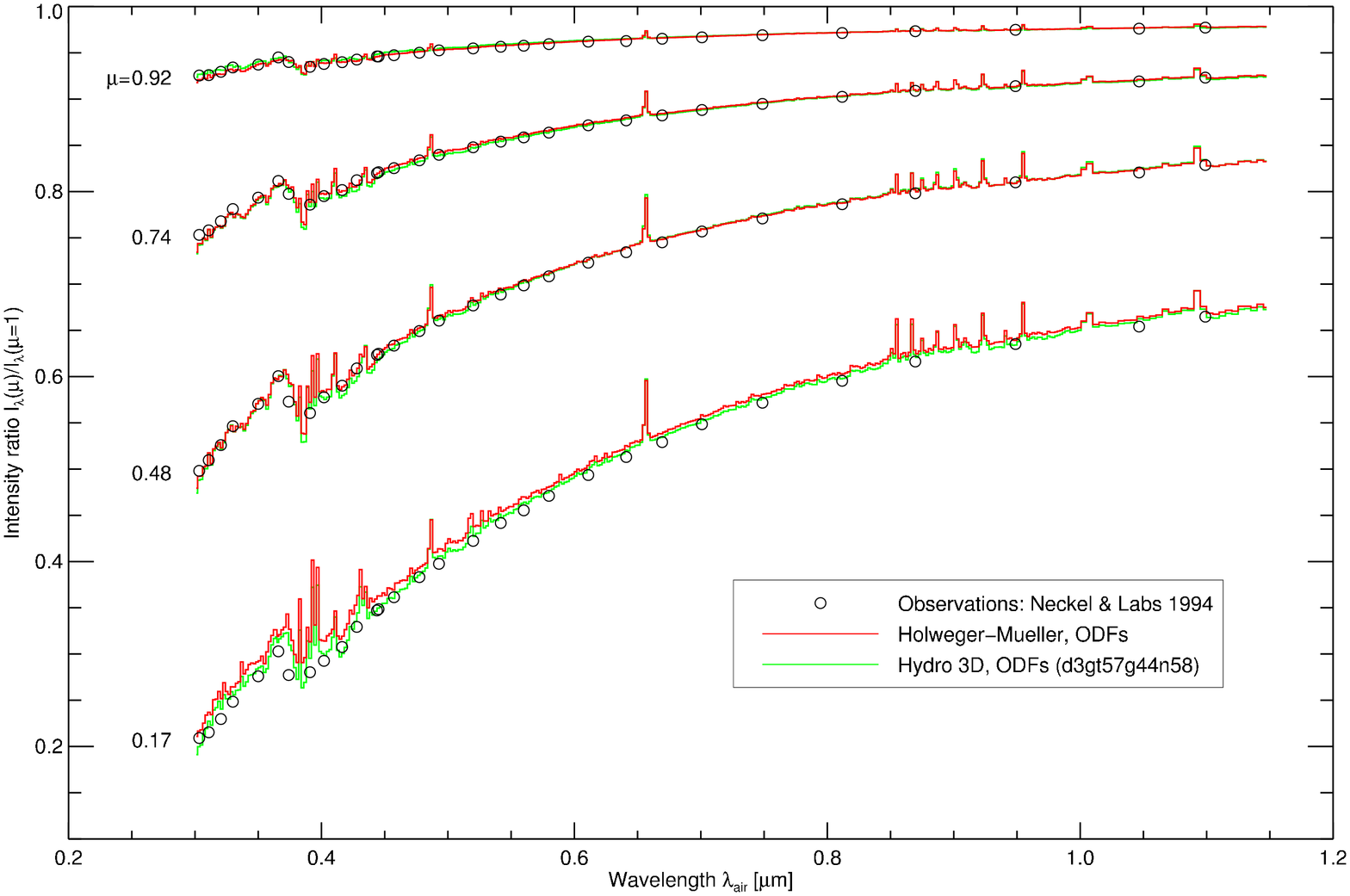} 
 \caption{Center-to-limb variation with line
     blocking using ATLAS ODFs: for four heliocentric angles the intensity
     relative to disk-center is depicted as a function of wavelength.} 
   \label{f1}
\end{center}
\end{figure}

In this section we want to demonstrate that 3D models atmosphere have reached
a high level of realism when it comes to the thermal structure of the lower
photosphere -- including temperature inhomogeneities due to granulation.
Figure~\ref{f1} illustrates the exceptional performance of our standard solar
\COBOLD\ model representing the center-to-limb variation of the solar
radiation field on a spatial scale where granulation is not resolved. The
calculation was done for a time series of 19 snapshots of the 3D flow field,
whose intensity pattern was subsequently horizontally and temporally averaged.
In the spectral synthesis calculations line blocking was accounted for by
applying an ATLAS (\cite[Kurucz 2005]{Kurucz05}) Opacity Distribution Function
with 1200 wavelength intervals, and 12 sub-intervals each. Fig. 1 shows the
emergent intensity averaged over the 12 sub-bins. The same calculation was
repeated for the 1D semi-empirical Holweger-M\"uller atmosphere
(\cite[Holweger \&\ M\"uller 1974]{hmsunmod}, HM). The overall match to the
observations by the 3D model is remarkable, including the wavelength range in
the Balmer continuum suffering from heavy line blocking. The precision is
challenging the available observations and the performance of the HM model
which was -- at least in part -- constructed to match the solar center-to-limb
variation.

\section{Disentangling the [OI]+Ni\,I feature at 630\,nm}

\begin{figure}
\begin{center}
 \includegraphics[width=0.8\textwidth]{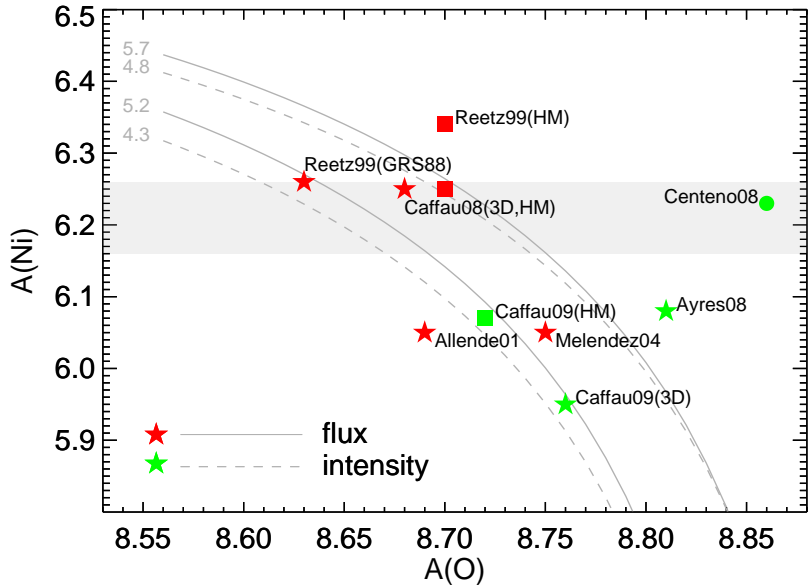} 
 \caption{Oxygen and nickel abundances obtained by various groups from the
   630\,nm feature. ``flux'' refers to disk-integrated, ``intensity'' to
   disk-center spectra. The solid and dashed curves delineate the relation
   between O and Ni abundance at fixed total equivalent width of the feature
   (labels in m\AA). The grey bar indicates the currently accepted range of the
   Ni abundance from other Ni lines. Further details see text.}
   \label{f2}
\end{center}
\end{figure}

The weak, forbidden oxygen line at 630\,nm which is intimately blended with an
even weaker line of neutral nickel, is considered as a prime abundance
indicator of oxygen in the solar atmosphere since the line is immune to
departures from LTE, and the blend lies in an otherwise rather clean part of
the spectrum. The oscillator strength of the transitions of O and Ni are well
determined so that one should expect that abundance determinations by various
groups should largely coincide. The only remaining difficulty should be the
separation of the total absorption in the feature into the contributions
related to O and Ni.  Figure~\ref{f2} summarizes the results obtained during
the last decade. All results have been normalized to the presently accepted
values of the oscillator strength of the O and Ni transition. The depicted
results were taken from: \cite{Reetz99}, \cite{ALA01}, \cite{Melendez04},
\cite{Ayres08}, \cite{Caffau08}, \cite{Centeno08}, and \cite{Caffau09}. Stars
indicate the application of theoretical model atmospheres in the analysis,
squares the HM model. Different from the others the work of Centeno et al. is
using spectro-polarimetric sunspot observations, and in this sense is
particular. The lines of constant total equivalent width of the O+Ni feature
were obtained with our spectral synthesis code and standard 3D solar model.

If all workers agreed in terms of model atmosphere and total equivalent
width of the feature all results should line up on a curve of constant
equivalent with in the O-Ni-abundance plane. However, even leaving aside the
result of Centeno et al. a large scatter has to be noted. There is a
noticeable influence of the applied model atmosphere, and also some effect of
the assumed equivalent width. Most strikingly perhaps, the separation into the
two components is far from unique. Here, the recent 3D based result of Caffau
et al. (2009) indicates a particularly low Ni abundance. While it is
difficult to reconcile with the presently accepted Ni abundance, it
provides a striking illustration of the still lingering problems in the
determination of solar photospheric abundances from spectroscopy.

\begin{acknowledgements}
\noindent\mbox{HGL, EC, BF, and PB acknowledge support from
EU contract MEXT-CT-2004-014265.}
\end{acknowledgements}

\end{document}